\documentclass[apj]{emulateapj}
\input psfig

\bibpunct{(}{)}{;}{a}{}{,}

\slugcomment{ApJ, in press}
\shorttitle{Long-Period Planets}
\shortauthors{Veras, Crepp, Ford}

\begin{document}
\title{Formation, Survival, and Detectability of Planets Beyond 100 AU}
\author{Dimitri Veras$^1$, Justin R. Crepp$^{1,2}$, Eric B.\ Ford$^1$}
\affil{$^1$Astronomy Department, University of Florida, 211 Bryant Space Sciences Center, Gainesville, FL 32111, USA}
\affil{$^2$California Institute of Technology, Department of Astronomy/Optical Observatories, 
1200 E. California Blvd., MC 105-24, Pasadena, CA 91125, USA}
\email{veras@astro.ufl.edu}
\begin{abstract}
Direct imaging searches have begun to detect planetary 
and brown dwarf companions and to place constraints 
on the presence of giant planets at large separations 
from their host star.  This work helps to motivate 
such planet searches by predicting a population of 
young giant planets that could be detectable by 
direct imaging campaigns.  Both the 
classical core accretion and the gravitational 
instability model for planet formation are 
hard-pressed to form long-period planets {\em in 
situ}.  Here, we show 
that dynamical instabilities among
planetary systems that originally formed 
multiple giant planets much closer to the 
host star could produce a population of giant 
planets at large ($\approx10^2$ AU $- 10^5$ AU) separations.
We estimate the limits within which these planets may survive,
quantify the efficiency of gravitational scattering
into both stable and unstable wide orbits, and
demonstrate that population analyses must take
into account the age of the system.  
We predict that planet scattering creates a 
population of detectable giant planets on wide 
orbits that decreases in number on timescales of $\sim 10$ Myr.
We demonstrate that several members of such populations
should be detectable with current technology,
quantify the prospects for future
instruments, and suggest how they could
place interesting constraints on 
planet formation models.
%
%
\end{abstract}


\section{Introduction}

Recently, several high-contrast imaging campaigns have begun to search
for young giant planets orbiting relatively nearby stars at large
orbital separations.  In addition to compelling detections of
likely planetary systems  
\citep[e.g.][]{kalas2008,maretal2008},
surveys of young stars have placed significant constraints on
the presence of any giant planets at large separations 
\citep[e.g.][]{mcczuc2004,kasetal2007,lafetal2007,nieetal2007}.  In the
near future, even more powerful searches have the potential to detect
even lower mass planets in wide orbits around nearby stars.  These
surveys could reveal new populations of extrasolar planets, since
planets with long ($\gtrsim 10^2$ yr) periods 
are difficult to detect by most
planet search techniques such as radial velocities, astrometry,
transit photometry, and even microlensing.

The frequency and orbital properties of long-period planets could provide 
interesting constraints for planet formation models.  
The core accretion model struggles even to explain the {\em in situ}
accretion of Uranus and Neptune, given the need to form a rocky core
while there is still significant gas in the circumsolar nebula
\citep{poletal1996,goldreich2004}.  This timescale
problem has led to several models in which Uranus and Neptune form
closer to the Sun and the migrate outwards to their present locations
\citep[e.g.][]{thoetal1999,thoetal2002,gometal2005,
moretal2005,tsietal2005,forcha2007}.  
Given these difficulties at 40 AU, the standard core
accretion model is unable to form giant planets at much larger
separations such as $\sim1000$ AU or even $\sim100$ AU. Alternatively, 
the gravitational instability model may become more plausible at
large separations \citep{rafikov2005,stawhi2008}.
Hence, some recent ground-based direct detection campaigns
have framed their searches as a way to search for planets that could
{\em not} be formed by core accretion, and thus provide support for
some planets having formed via gravitational instability.  However,
recent research has also cast doubt on the viability of the
gravitational instability model 
\citep{mejiaetal2005,caietal2005,boss2006,rafikov2007}.
Therefore, one might worry that a non-detection by direct 
imaging planet searches 
\citep[e.g.][]{kasetal2007,lafetal2007,nieetal2007}
would fail to provide new constraints on planet formation models.  
In this paper, we show that this is not the case.  In either the 
core accretion or
gravitational instability model, the formation of multiple giant
planets at moderate separations ($\sim1-10$ AU) naturally leads to
dynamical instabilities that result in one or more planets being
scattered into very wide ($\approx10^2$ AU $- 10^5$ AU) orbits.
Thus, direct imaging searches (such as GPI -- The Gemini Planet Imager, 
\citealt*{macetal2006a}) and microlensing planet searches have
the potential to place significant constraints on the frequency of
dynamically active planetary systems that form multiple giant planets.

\label{SecFormation}
Extrasolar giant planets have a broad range of eccentricities.  
Although several proposed mechanisms may explain
the eccentricities of some planets, the high frequency of eccentric
orbits among giant planets suggests that dynamical instabilities
resulting in planet-planet scattering play an important role in
sculpting many planetary systems.  Although many variations of the
planet-planet scattering model of been explored 
\citep[e.g.][]{rasfor1996,weimar1996,levetal1998,foretal2001,
marwei2002,adalau2003,verarm2004,verarm2006,mooada2005,
chaetal2008,forras2008,jurtre2008},
these studies focused on the final properties
of the bound planets and paid little attention to the orbital
evolution of planets that will eventually be ejected from the system.
Recently, a dynamical instability has been suggested as a possible
formation mechanism for the substellar companion to GQ Lup 
\citep{debsig2006} in a wide ($\approx 100$ AU) orbit.  However, 
previous dynamical
studies have yet to quantify the probability for scattered planets to
be detected via direct imaging.  We investigate the orbital properties
and lifetime of planets scattered into wide orbits and assess the
prospects for their detection via high-contrast imaging. Our study
complements that of \cite{schmen2008}, as we consider
additional effects, including galactic tides, scattering of
terrestrial-mass planets, and observational estimates of the contrast
for long-period giant planets.

Given the history of extrasolar planets repeatedly being discovered in
unexpected locations, it is worth searching for planets anywhere that
they could survive.  Therefore, we begin by exploring the physical limits that
can be placed on very long-period planets and find that planets could
survive with orbital separations of up to $\sim10^5$ AU in \S\ref{anal}.
Such estimates can be compared to the separations at which planets are 
observed in order to determine whether there are additional processes shaping
this distribution.
In \S\ref{dimi}, we describe a model where multiple giant planets form at
traditional orbital separations ($\sim3-30$ AU).  As the disk
dissipates, the planets will perturb each other, eventually resulting
in eccentricity growth, close encounters, and some planets being
ejected from the planetary system \citep{chaetal2008}.  Before
the planets are ejected, they will spend considerable time at very
large orbital separations from the host star. Thus, planet scattering
should result in a significant number of giant planets being scattered
onto very wide orbits, where they could be detected by recent or
future direct imaging planet searches.  By quantifing the 
efficiency of scattering onto such orbits, we estimate how large
a sample needs to be surveyed in order to detect such planets.
We additionally explore the timescale on which detectable wide-orbit planets
remain; we predict that the population of unstable planets
``passing through'' a wide-orbit phase is larger than the 
population of planets on stable wide orbits.
The existence of this
population does not require that either the standard core accretion or
gravitational instability mechanism be able to form giant planets at
large separations, where direct imaging campaigns are most sensitive.
Observations of these planets at different ages can impart information
about the chaotic stage of planet formation (e.g., how many giant
planets are typically formed before planet scattering widdles down
the number of surviving planets to 2 or 3).
Finally, we consider the prospects for a direct imaging
campaign to detect such planets and their implications in \S\ref{observe}.
System age, contrast, and separation limits all affect population
analyses from such campaigns.

\section{Survival and Destruction of Long Period Planets}\label{anal}

We begin by asking what processes
would limit the survival of very long-period giant planets orbiting
main-sequence stars.  Most previous studies investigating the effects
of passing stars have focused on planets at separations $\le$ 40 AU
\citep[e.g.][]{lauada1998,foretal2000a,adaetal2006}.  If
giant planets have much larger orbits, then the effects of stellar
encounters could be dramatically increased.  In the limit that planets
are much less massive than stars, they can be treated as test
particles, allowing us to apply many arguments previously applied to
the survival of long-period comets.

\subsection{Field Stars}
\label{SecFieldStars}

First, we consider the constraints on long period planets around field
stars in the galactic disk, closely following the treatment of
\cite{tremaine1993}.

\subsubsection{Galactic Tide}

Planets in sufficiently wide orbits will be influenced by
gravitational perturbations from the galactic tides.  Like comets, a
planet will typically become unbound if the star-planet separation
exceeds the tidal radius, estimated by
\begin{equation}
a_t \simeq 1.7\times10^5 \mathrm{AU} \left(\frac{M_\star}{M_\odot}\right)^{1/3} \left(\frac{\rho}{0.1M_\odot \mathrm{pc}^{-3}}\right)^{-1/3},
\label{at}
\end{equation}
where $M_\star$ is the mass of the host star, and $\rho$ is the local
galactic mass density \citep{tremaine1993}.  Once a planet is scattered beyond
$a_t$, it will likely become unbound on an orbital timescale.  

\subsubsection{Passing Stars}

Encounters with passing stars and giant molecular clouds can eject a
planet, either by a single encounter or by repeatedly perturbing the
planet's orbit.  \cite{weietal1987} calculate that half-life of a
long-period planet in the solar neighborhood due to the cumulative
effect of many encounters with passing stars to be
\begin{equation}
\tau_{\rm ps} \simeq 10 \mathrm{Gyr} \left(\frac{M_\star}{M_\odot}\right) \left(\frac{10^4 \mathrm{AU}}{a}\right) \left(\frac{0.15M_\odot \mathrm{pc}^{-3}}{\rho}\right), 
\label{time}
\end{equation}
where $a$ is the planet's semi-major axis.  
Assuming that the planet either formed {\em in situ} 
or migrated to its current location on a timescale 
much less than the system age,
we can equate the timescale for a planet to become
unbound to the age of the star, resulting in an estimate of the maximum
distance at which a planet is likely to survive, 
\begin{equation}
a_{\mathrm{ps}} \simeq 10^5 \mathrm{AU} \left(\frac{\mathrm{Gyr}}{t_\star}\right) \left(\frac{M_\star}{M_\odot}\right) \left(\frac{0.15 M_\odot \mathrm{pc}^{-3}}{\rho}\right),
\label{aps}
\end{equation}
where $t_\star$ is the age of the star, and provided that
$a_{\mathrm{ps}}\ll a_t$. 
Thus, the limiting semi-major axis for
young disk stars will be set by the tidal limit, while the limiting
semi-major axis for old field stars will be set by the cumulative 
effect of many stellar encounters.

\subsection{Stars in Clusters} \label{SecClusters}

Most disk stars formed in small clusters 
\citep{thretal2001}.  Because the rate of stellar encounters 
depends strongly on the
local stellar density, the effects of stellar encounters are likely 
dominated by interactions while the star is young and still in a dense
young stellar cluster.  If planets form or migrate to long period
orbits while the host star is still in a dense region of the cluster,
then encounters with cluster members could place more stringent
constraints than those discussed in \S\ref{SecFieldStars}.  The
importance of stellar encounters is linked to the dynamical evolution
of the cluster in which the star formed.  Following \cite{adaetal2006}, 
we separate the dynamical evolution of the cluster from the
dynamics of close encounters.

\subsubsection{Cross Sections for Impulsive Encounter with Single Stars}

First, we consider the ejection of planets 
in a single encounter.  
The timescale for a stellar encounter to eject a planet from 
its host star can be estimated by $\tau^{-1} \sim 
\left< n_\star \sigma v_{\infty} \right>$, where $n_\star$ 
is the local density of stars, $v_{\infty}$ is the 
velocity at infinity of the passing star, $\sigma$ is 
the relevant cross section, and $\left< \right>$ indicates a time average.  
For very long-period planets, encounters with single stars are
impulsive when
\begin{equation}
a>a_{\rm imp} \simeq 890 \mathrm{AU} \left(\frac{M_1+m}{M_\odot}\right) 
\left(\frac{\mathrm{1 \ km/s}}{v_\infty}\right)^2,  
\label{EqnImpulsive}
\end{equation}
where $a$ is the semi-major axis, and $M_1$ is the mass of the host star.
In this regime, the cross section for ionizing the planet from the host
star is
\begin{equation}
\sigma_{\rm ion}\simeq \frac{40\pi a G}{3v_{\infty}^2} \frac{M_2^2}{M_1+m} \simeq (193 \mathrm{AU})^2 \left(\frac{a}{AU}\right) \left(\frac{\mathrm{km/s}}{v_\infty^2}\right)^2 \frac{M_2^2}{M_1}, 
\end{equation}
where $M_2$ is the mass of the passing star, and $G$ is the
gravitational constant 
(\citealp*{hut1983};Eq.\ 13 of \citealp*{freetal2006})

In addition to planets being ejected, some planets will be exchanged
into orbit around the passing star.  For the sake of comparing to 
the cross section for ionization, we
consider equal mass stars of mass $M=M_1=M_2$, in which case
\begin{equation}
\sigma_{\rm ex_s(nonres)} \simeq \frac{20\pi}{3 a v_\infty^6}G^3 M^2 (M+m)
\end{equation}
(\citealp*{hut1983};Eq.\ 14 of \citealp*{freetal2006}).  For 
strongly impulsive
encounters, the cross section for exchange is much less than the cross
section for ionization, so exchanges are relatively insignificant.

\subsubsection{Cross Section for Intermediate Encounter with Single Stars}

For planets at somewhat shorter orbital periods ($a\le1000$ AU), the
scattering behavior enters an alternative regime, where the cross
sections for ionization and exchange are comparable.  Numerical
simulations for a circular planet encountering a single star find that
\begin{equation}
\sigma_{\rm ion} \simeq 1.2 \pi a G M_t / v_\infty^2,
\end{equation}
and
\begin{equation}
\sigma_{\rm ex_s(nonres)} \simeq 0.6 \pi a G M_t / v_\infty^2,
\end{equation}
were $M_t=M_1+M_2+m$ is the total mass 
\citep[Eq. 15 and Fig. 3 of][]{freetal2006}.  
In this regime, 
when the planet is exchanged to
the the passing star, the binding energy of the planet before and
after the star will be comparable (with a typical change of
$\simeq20\%$).  Therefore, exchanges between stars of similar mass
result in a planet orbit comparable to before the encounter.  Among a
coeval cluster of stars, such an exchange is unlikely to be
recognizable, except in unusual circumstances 
\citep[e.g., PSR 1620+26;][]{foretal2000b}.  

\subsubsection{Rate of Close Encounters for Stars in Clusters}

Simulations of star clusters can be used to calculate the rate
of stellar flybys.  Table 3 from \cite{adaetal2006} provides empirical
fits to the results of their simulations, with the form,
\begin{equation}
\Gamma = \Gamma_o \left(\frac{b}{1000\mathrm{AU}}\right)^\gamma,
\end{equation}
where $b$ is the closest approach distance, and $\Gamma_o$ and
$\gamma$ are fit parameters \citep[see Table 3 of ][]{adaetal2006}.
For a typical cluster of 100-1000 stars, they find that a typical star
undergoes $N(<100 \ \mathrm{AU}) \sim 0.2-2$ or interactions with
closest approach distance less than 100 AU over the course of a 10 Myr
simulation.  Similarly, they find $N(<1000 \ \mathrm{AU}) \sim 1-14$
encounters coming within 1000 AU over the course of a 10 Myr
simulation.  

\subsubsection{Ionization Rate by Single Encounters with Single Stars}

We find the semi-major axis beyond which a planet will typically 
be ejected by a single encounter with a passing star by 
setting $b=\sqrt{\sigma_{\rm ion}/\pi}$.  Thus, a typical 
planet will be ejected if the encounter is impulsive (i.e.,
$a>a_{imp}$) and the planet's orbit exceeds
\begin{equation}
a_{ps,imp} \simeq 85 \mathrm{AU} \left(\frac{(M_1+m)M_\odot}{M_2^2}\right) \left(\frac{v_\infty}{\mathrm{km/s}}\right)^2 \left(\frac{\mathrm{Myr}}{\Gamma_o t}\right)^{2/\gamma},
\label{EqnApsImp}
\end{equation}
where $t$ is the time avaliable for stellar encounters.  Similarly, a
typical planet will be ionized if the encounters are in the
intermediate regime identified by \cite{freetal2006} and
\begin{equation}
a_{ps,int} \simeq 944 \mathrm{AU} \left(\frac{M_\odot}{M_t}\right) \left(\frac{v_\infty}{\mathrm{km/s}}\right)^2 \left(\frac{\mathrm{Myr}}{\Gamma_o t}\right)^{2/\gamma}.
\label{EqnApsInt}
\end{equation}

For planets that form concurrently with the star and {\em in situ},
the final factor of Eqs. (\ref{EqnApsImp}) \& (\ref{EqnApsInt}) ranges
from $\simeq0.5-6.9$ for the 10 Myr simulations of small clusters with
100-1000 stars \citep{adaetal2006}.  Therefore, planets that form (or
migrate) beyond $a_{imp}$ during the earliest stages of the cluster
would be ejected by a single impulsive stellar encounter.  For some
choices of cluster parameters (i.e., the ``cold start'' simulations of
\citealp*{adaetal2006}, planets inward of $a_{imp}$ but beyond $a_{ps,int}$
could be ionized by passing stars before the cluster dissolves.

The previous estimates assume that the planet forms {\em in situ} at
the same time as the star.  As we discussed in \S\ref{SecFormation}, 
some theoretical models
predict that planets might form close to a star and then be scattered
outwards much later.  For the same simulations, but considering
planets that are placed at large separation 5 Myr into the cluster
simulations, the final factor of Eqs. (\ref{EqnApsImp}) \&
(\ref{EqnApsInt}) ranges from $\simeq4.6-190$ (considering the
subsequent 5 Myr evolution of the clusters; \cite{adaetal2006}.
Therefore, planets that migrate to
large separations after the cluster loses its gas 
\citep[at 5 Myr in simulations of][]{adaetal2006} could avoid being 
disrupted by stellar encounters at separations of up to
$\sim\mathrm{max}(a_{imp},a_{ps,imp})\sim 10^3-3\times 10^4$ AU.

Most stars in a cluster are actually members of binary (or higher
order multiple) star systems.  For the wide planetary orbits of interest,
encounters with close binaries proceed similarly to encounters with
single stars.  The primary difference is that encounters which
previously resulted in the planet exchanging into orbit around the
passing star now result in forming a triple system including the
circumbinary planet.  Encounters with binary stars that have larger
orbital periods are more likely to result in ionization, as even
exchange interactions often result in unstable triple systems that
subsequently eject the lowest mass planet.  Numerical simulations 
including a plausible distribution of binary mass ratios and 
periods verify that the cross section for ionization by binary 
systems is roughly comparable to the sum of the cross sections 
for ionization by the individual stars.

%

\subsection{Summary of Survival of Long Period Planets}
In summary, interactions prior to the dissociation of the cluster can
dominate the ionization of planets that form or migrate to large
distances before the parent cluster dissociates.  For planets that
appear at large separations near the birth of the cluster, survival
for the life of the cluster can require $a\le500-6000$ AU, depending on
the cluster properties.  However, if planets are able to form or
migrate to very large separations more than a few million years after
the formation of the cluster, then they could survive at much larger
distances.  A delay of 10 Myr is more than sufficient to allow planets
to survive with semi-major axes approaching the tidal limit (or
$a_{\mathrm ps}$ for old star).

In order to identify the most significant processes affecting planets 
in wide orbits, we compare several relevant timescales.  The planet 
formation timescale is a key difference among the two leading models 
for planet formation: core accretion and gravitational instability 
\citep[see][ for a recent summary]{armitage2007}.  
In the classical core accretion model, accretion of gas onto a giant 
planets at Jovian-like distances ($\sim 5$ AU) lasts at least 
$8$ Myr \citep{poletal1996}.  However, \cite{haietal2001} estimates that 
the lifetimes of protostellar disks last at most $\approx 3$ Myr, and 
this timescale imposes a maximum value on the core accretion timescale.
This has inspired theoretical research to explore whether core accretion 
may be able to act more rapidly.  Indeed, variations on the core accretion 
model are able to form planets more rapidly.  For example, \cite{hubetal2005} find 
that Jupiter could have formed in $2.2$ Myr, and \cite{dodetal2008} present two 
new models where Saturn forms at $\sim12$ AU in just under $3.5$ Myr.  
\cite{chambers2006} envisages accretion times under 1 Myr.  While further 
research is in order, it is unlikely that the core accretion timescale lies 
outside of $0.1-10$ Myr.  In the gravitational instability model, giant 
planets acquire their mass on much shorter timescales which are unlikely 
to surpass $10^4$ yr \citep[][ and references therein]{duretal2007}.

After multiple planets have formed in a protoplanetary disk, they can begin to 
interact with each other.  Initially, the gas disk can suppress eccentricity 
excitation, but once the gas dissipates, the system can rapidly become orbit 
crossing.  Alternatively, for systems containing planets that are more widely 
spaced, the planets can remain on low-eccentricity orbits for over a billion 
years before chaotic evolution leads to orbit crossing.  In either case, 
once orbits cross, close encounters and eccentric orbits soon follow 
\citep{chaetal2008}. Before a planet is to be scattered onto a very wide orbit 
or ejected from the system, it will undergo a series of many close encounters.  
The timescale depends on the planet masses and the orbital period of the more 
massive planet (that typically remains bound).  For scattering by Jupiter, 
the median value of this timescale is $\sim 10^4$ yr, with a range that 
spans several magnitudes (see \S\ref{results}).

The process which limits the orbital separation of long-period planets will vary 
from system to system, depending on the host star environment and the time 
until the planet is scattered onto a wide orbit.  Most stars are believed to 
form in star clusters, groups, or associations that dissociate on timescales of 
a few Myr.
Simulations of star clusters show that most 
of the close encounters occur during the first few million years.  If planets form 
{\em in situ} in wide orbits (e.g., by gravitational instability) or if they
migrate or are scattered into wide orbits very rapidly, then planets 
beyond $\sim 700-4000$ AU 
could have been disrupted by passing stars before the host cluster dissociates 
\citep{adaetal2006}.  The width of the main sequence in open clusters suggests 
that stars in a larger cluster form within $\sim~5$ Myr of each other 
\citep{bonbic2009}. 
Thus, it may be possible for planets formed via gravitational instability to 
survive at larger separations, if the planet orbits a star that formed a few 
Myr after the cluster's massive stars have started to remove the cluster's gas.  
For such planets and planets that are scattered into wide orbits shortly after 
the cluster dissociates, the tidal field of the molecular cloud can disrupt 
planets with separations exceeding $\sim~10^4-10^5$ AU.  For planets that are 
scattered into wide orbits well after the cluster has dissociated and the 
star has drifted away from the host molecular cloud, perturbations from passing 
stars and galactic tides can remove wide-period planets on timescales of 
$\sim$ Gyr \citep[Eq. \ref{time};][]{higetal2007}.

For direct imaging to detect planetary systems like our solar system, it is important 
to focus on the closest young stars, where the inner working angle corresponds to a 
small physical separation.  Most young stars in the immediate solar neighborhood are 
believed to have formed in small groups or associations, in which case the processes 
described in \S~\ref{SecFieldStars} will not be relevant.  However, 
these same direct imaging 
campaigns may overlook young giant planets on wide orbits, as they lie beyond the 
detector's outer working angle.  When searching for planets at wide separations, it 
is possible or even advantageous to target more distant star forming regions.  Thus, 
it would be possible to search for planets that formed in more substantial 
clusters that span a variety of ages.

\section{Formation of Long Period Planets Via Planet-Planet Scattering}\label{dimi}
\label{setup}
\label{SecScattering}
When considering the survival of long-period planets in
\S~\ref{SecFieldStars}, we applied arguments previously developed for
the survival of Oort cloud comets.  However, the mechanisms
responsible for creating the Oort cloud are unlikely to produce a
large population of long-period planets.  Populating the Oort cloud is
inefficient, as only $\sim$2\% of the parent population
scattered by Jupiter and Saturn end up in bound orbits with very long
periods, rather than being ejected from the system or colliding with a
planet \citep{dunetal1987,donetal2004,higetal2006,braetal2007}.  
In our solar system, Uranus and Neptune may
scatter planetesimals into the Oort cloud more efficiently, largely
due to their lower masses.  However, they are not massive enough to
scatter a Jupiter-mass planet into an Oort cloud-like orbit.  Assuming
that planet formation typically produces a few giant planets per star,
a typical cluster might form a few systems where a giant planet is
scattered outwards and be perturbed onto an Oort cloud-like orbit that
will persist for billions of years.  Although giant planets remaining in
very long-period orbits for billions of years will be rare, the number
of giant planets at large separation may be greatly increased around
young stars, since giant planets that have been scattered outwards
will typically spend less than $1$ Myr on wide, eccentric
orbits before becoming unbound.  The chance of detecting such planets
is further enhanced since the direct detection of giant planets is
most practical while they are still young and radiating the heat of
accretion in the infrared.  Therefore, we explore the possibility of
detecting giant planets that have recently been scattered outwards,
before they eventually become unbound.

\subsection{Methods}

\subsubsection{Initial Conditions}

We investigate whether planet scattering may produce a population of
detectable planets on wide orbits by performing n-body integrations of
``dynamically active'' (= will undergo instability) planetary 
systems that typically contain three giant planets and three 
less massive planets.
In the test particle limit of the less massive planets,
systems with three giant planets represent simple
systems that can undergo a dynamical instability after an extended
period of seemingly regular orbital evolution 
\citep{marwei2002,chaetal2008}.  This is in contrast to
systems containing only two planets, which have a sharp 
boundary dividing
provably stable configurations from systems that rapidly result in
close encounters \citep{gladman1993}.  Three-planet systems have the
additional advantage that they are more likely to retain signatures of
possible prior instabilities \citep{verarm2006}.  Although planetary
systems may form even more planets, chaotic orbital evolution
typically results in planets being ejected, accreting onto the star,
or colliding with each other until only 2 or 3 planets remain
\citep{papter2001,adalau2003}.  Due to the strongly chaotic evolution
of dynamically active multiple-planet systems, systems with three or
more giant planets behave qualitatively similar \citep{jurtre2008}.  

We consider a realistic ensemble of post-formation initial stellar
masses, planet masses, and orbital separations.  
We draw stellar masses from two different initial mass 
functions, and show that the results are broadly similar.
We first utilize a simplistic initial mass function 
(IMF) with a power-law
exponent of $-1.35$ with mass cut-offs at $0.3 M_{\odot}$ and $3.0
M_{\odot}$, approximating the distribution of A through mid-M
stars.  We then apply rejection sampling in order to obtain
a more realistic Miller-Scalo \citep{milsca1979} power-law 
distribution (with exponents of $-0.25$, $-1.0$ and $-1.3$ for 
the mass ranges 
$0.3 M_{\odot} - 1.0 M_{\odot}$,  $1.0 M_{\odot} - 2.0 M_{\odot}$ and
$2.0 M_{\odot} - 3.0 M_{\odot}$, respectively).  We report all data
and figures from the Miller-Scalo power-law distribution, and where
comparison is appropriate, data for the simple power-law IMF 
in square brackets.  The comparisons will show that the results
are insensitive to this choice of IMF at the $\sim 1\%$ level.

We randomly draw six planet masses from a distribution uniform
in the log of the planet-star mass ratio 
as suggested by radial velocity surveys 
\citep{cumetal2008}.  The upper limit ($10^{-2}$) is based on the paucity
of more massive planets around solar-type stars.  The lower limit 
($9 \times 10^{-6}$) is near a terrestrial mass and is well-beyond 
the reach of current direct imaging campaigns.  

%

Theoretically, the site of giant planet formation is believed to occur
predominantly beyond the location of the snow line.  Therefore, we 
randomly sample the innermost planet's semi-major axis uniformly 
from $3-7$ AU.  The lower bound roughly corresponds to the location 
of the snow line in the minimum mass solar nebula \citep{lecetal2006}.  
Various environments can yield snow line locations which vary over one order
of magnitude \citep{garlin2007,kenken2008}.  Therefore, our range of
semi-major axes for the innermost planet allows for a variety of likely
snow line locations.  

We choose an initial spacing between neighboring planets constant in
units of the mutual hill radius, $K$.  Our choice of $K=4.2$ 
almost guranatees that each system will undergo dynamical 
instability, and does so within a few million
inner orbital times (see Fig. 29 of \citealt{chaetal2008}).  Thus,
instabilities occur only after enough orbits that chaos has erased
memory of the exact initial conditions. 
The value of 
$K$ then effecively sets the timescale for instability in systems
where we expect close encounters, collisions and ejections to 
occur.
Hence, our results should be interpreted in terms of the 
rate of detectable planets per {\it dynamically active} planetary 
system, rather than the rate relative to all stars or all 
planetary systems.
Although our choice for the value of $K$ sets the 
timescale for the initial onset of instability in our 
simulations, the final outcomes and the timescale for 
planets on wide, eccentric orbits to become unbound is not 
sensitive to $K$, but rather the mass of the interior planet 
that eventually ejects the planet \citep{pansar2004}.

We randomly choose the orbital 
orientation and mean anomalies from uniform
distributions, and assign the initial eccentricities and inclinations
also from uniform distributions, bounded by $<0.01$ and $<0.01^{\circ}$.
Because we begin each system with three relatively closely spaced massive
planets, the systems will typically self-excite larger eccentricities and
inclinations.  In this paper, we focus our attention on planets that have
undergone repeated close encounters with one or more planets before
being scattered to very large ($\gtrsim 100$ AU) separations.  
Thus, our conclusions will
not be sensitive to the exact choice of initial semi-major axes,
eccentricities, and inclinations.

\subsubsection{Integration Methods}

We integrate 1899 [2000] sets of initial conditions described above 
for 500 million years.  We perform all integrations with 
the Burlirsch-Stoer
integrator from the {\it Mercury N}-body integration package
\citep{chambers1999}.  
In order to monitor the effects of ejected planets, planets
were removed from the system only upon reaching a semimajor
axis of $10^6$ AU from their parent star.
All integrations included the effects of
possible collisions of planets with one another and with the parent
star, but did not include fragmentation.  A different set 
of 2000 integrations also
included tidal effects from the galactic disk added to the function
{\it mfo\_user}, according to the prescription given by
\cite{higetal2007}:

\begin{equation}
\frac{d^2 \vec{r}}{dt^2} = 
-G M_{\odot} \frac{\vec{r}}{r^3} 
-\nu_{0}^2 \vec{z}
\label{galtide}
\end{equation}

\noindent{where} $\nu_{0} = \sqrt{4 \pi G \rho}$ is the vertical
frequency, and $\rho$ is the density in the solar neighborhood.  We
assumed $\rho = 0.1 M_{\odot} {\rm pc}^{-3}$ \citep{holfly2000}.  In
addition to the disk tide, there is a radial component of the galactic
tide due to the galactic bulge and dependent on the Sun's angular
speed around the galactic center.  Since the radial galactic tide
operates on a time scale an order of magnitude longer than the disk
tide \cite{higetal2007}, we neglect this term.

In \S\ref{observe}, we will
discuss the prospects for observing such planets.  In this section, we
focus on the dynamical properties of these integrations.

\subsection{Results} \label{results}

The numerical simulations described in \S\ref{setup} result in
several qualitative outcomes.  In each case, at least one of the
six planets is ejected, collides with its parent star, or collides
with one another.  Any of these outcomes causes the planet
to be removed from the system.  
The following lists the percent
of systems with a given number of bound planets remaining at the
end of the simulations (at $500$ Myr), with the values 
in parentheses indicating
the simulations which included the galactic tide:  
0 planets, $1.4$\% [$1.2$\% ($1.2$\%)];
1 planet, $15.4$\% [$15.6$\% ($15.5$\%)];
2 planets, $41.0$\% [$41.3$\% ($41.4$\%)];
3 planets, $27.9$\% [$27.9$\% ($27.7$\%)];
4 planets, $13.3$\% [$13.0$\% ($12.2$\%)];
5 planets, $1.16$\% [$1.15$\% ($1.95$\%)];
6 planets,  $0$\% [$0$\% ($0$\%)].
Hence, the broad dynamical effect of the galactic tide on 
these simulations, after $500$ Myr and over 
distances of $10^6$ AU, is negligible.  Further, 
two-planet systems are the most likely outcome of our planet-planet
scattering simulations, with three-planet systems being the
second-most likely outcome; these results agree well with 
Fig. 6 of Adams \& Laughlin (2003), who use 10 planets
and a similar initial mass distribution.


\begin{figure}
\plotone{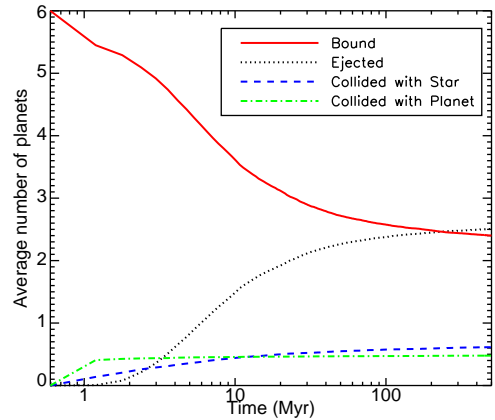}
\figcaption{Average number of planets vs. time for planets
which: remain bound to their parent star (red/solid line), have been ejected
(black/dotted), have collided with the star (blue/dashed line), and have collided with another planet (green/dot-dashed line).
\label{frem}}
\end{figure}

Although our simulations were run for 500 Myr, the vast majority
of close encounters, collisions and ejections occur well before
100 Myr. The {\it median} time to the first close encounter between any
two planets or planet is $t \approx 9.04 \times 10^3 \ [9.15 \times 10^3]$ yr,
and the {\it mean} time to the first close encounter was 
$t \approx 1.33 \times 10^6 \ [1.30 \times 10^6]$ yr. We take one Hill
Radii to be the maximum distance from a body which constitutes
a close encounter.  In order to help quantify the dynamical settling time
of the systems, we plot the average number of planets remaining,
planets ejected and planets suffering a collision as a function
of time in Fig. \ref{frem}.  The figure demonstrates that approximately
half of the planets in each system still remain after $2 \times 10^7$ yr,
and that ejections (as opposed to collisions) become the predominant 
removal mechanism after $30$ Myr.  
After $2 \times 10^8$ yr, on average, more planets
have been ejected than remain. Further, the fraction of planet-planet
collisions are approximately equal to the fraction of planet-star
collisions, but the latter dominates after $1 \times 10^7$ yr.  At the
end of our simulations, on average we can expect a system to still
have between 2-3 planets.  On even longer timescales we suspect
that just a few more planets might be removed from the systems.

\begin{figure}
\plotone{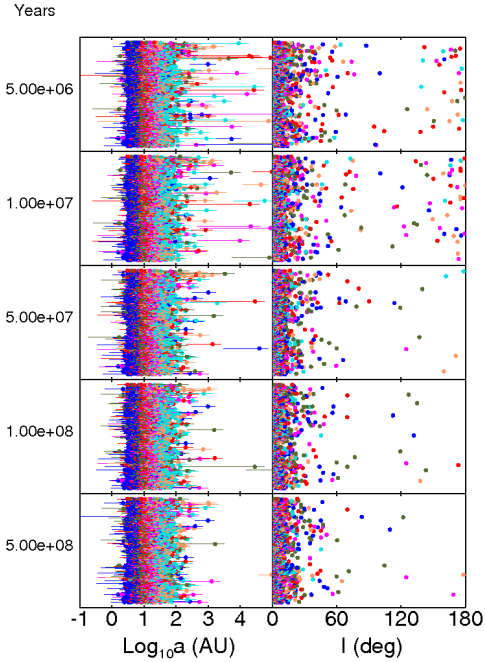}
\figcaption{
Snapshots showing the orbital properties of the surviving planets (of 
each system) with masses greater than $5 \times 10^{-4} M_{\odot}$
at five times from 5 Myr (top) to 500 Myr (bottom). 
Within a panel, each row represents a different system. 
In the left panels, points indicate
semimajor axes, and the horizontal lines 
indicate the pericenter and apocenter distance for 
each planet.
In the right panels, points display inclination.
The planet symbol colors indicate indicate initial 
proximity to the star, and are given by
blue/circles (innermost), green/crosses, 
red/squares, magenta/triangles, salmon/diamonds 
and aqua/dots (outermost).   
\label{fsnap}}
\end{figure}

Figure {\ref{fsnap}} presents snapshots of the semi-major axes,
periastron distances, and apoastron distances for each of the
bound planets remaining with masses greater than 
$5 \times 10^{-4} M_{\odot}$ (a representative observational constraint) 
at times of 5, 10, 50, 100 and 500 Myr.
Each row represents a different system, and the five panels, in descending
order, contain 3583, 3476, 3227, 3164 and 3056 
[3737, 3623, 3364, 3298, and 3186] planets (from an initial
population of 12,000 planets).  For the simple power law, without the
mass constraint, the total number of bound planets at each of those 5 times
would be [7537, 6805, 5403, 5122 and 4783].  
The horizontal lines indicate the periastron-apoastron 
range of each planet, and the points indicate the semi-major axes.
Color indicates the initial ordering in terms of semi-major axis, 
as described in the caption.  Because planets that have 
been scattered beyond 100 AU are typically the least
massive planets in the system, and there does not appear to be a significant
correlation with the initial ordering (in terms of semi-major axes).
By 5 and 10 Myr, the planet scattering has significantly excited
planetary inclinations, but inclinations of $\approx 90^{\circ}
\lesssim i \lesssim 150^{\circ}$ are rarer than those
of $\approx 150^{\circ}
\lesssim i \lesssim 180^{\circ}.$
We find $0.5\%$ [$0.5\%$] of all planets at 5 Myr both harbor 
semi-major axes above 100 AU and are on retrograde orbits, 
but these planets are typically ejected after several $10^7$ years.

\begin{figure}

\plotone{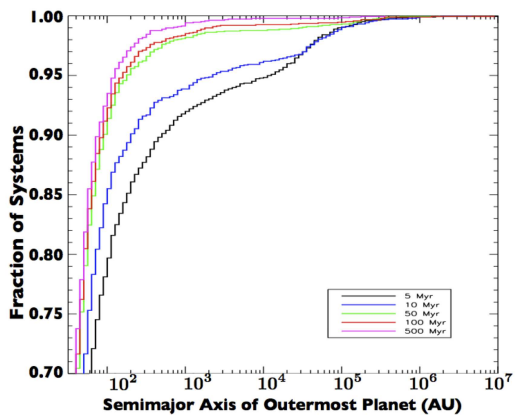}
\figcaption{
Cumulative histograms of the semimajor axis of the outermost bound planet
at 5 Myr (black/solid lines), 10 Myr (blue/dotted lines), 
50 Myr (green/dashed lines), 100 Myr (red/dot-dashed lines)
and 500 Myr (magenta/dot-dot-dot-dashed lines).
\label{fsemi}}
\end{figure}

The planets of greatest interest to observational searches are those
on ``wide'' orbits, with semimajor axes of tens, hundreds or thousands
of AU.  The number of planets surviving beyond 1000 AU declines rapidly
from 10 Myr to 50 Myr, but even at 500 Myr,
planets remain beyond 1000 AU.
In Fig. \ref{fsemi}, we present a
cumulative distribution for the semimajor axis of the
outermost surviving planet from $10^{1.5}-10^7$ AU at 5 Myr (black/solid lines),
10 Myr (blue/dotted lines), 50 Myr (green/dashed lines), 
100 Myr (red/dot-dashed lines) and 
500 Myr (magenta/dot-dot-dot-dashed lines).  The ejection 
of many planets (especially between 10 Myr and
50 Myr) partially accounts for the difference between the two sets of
curves (compare the 5 and 10 Myr curves to the 50, 100 and 
500 Myr curves).
Fewer planets reside beyond 40 AU at 5 Myr than at 10 Myr, because
many systems are still undergoing significant dynamical evolution at
10 Myr.  Between 10 Myr and 100 Myr, the number of planets on wide
orbits decreases, since the rate of planets being scattered
outwards into wide orbits is less than the rate of planets already on
wide orbits becoming unbound.  At 100 Myr, less than $0.1\%$ of
systems harbor planets with semi-major axes beyond $10^5$ AU.  Although
small, such planets could be present in a cluster of thousands of 
stars.

\begin{figure}

\plotone{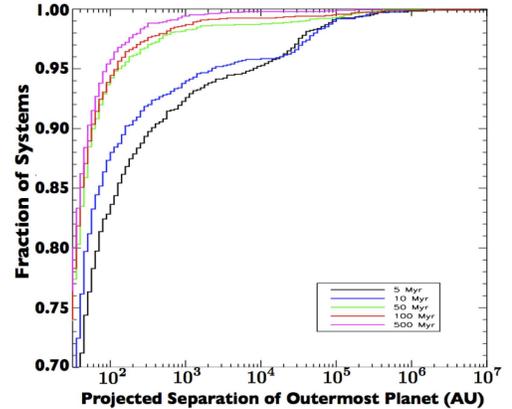}
\figcaption{
Cumulative histograms of the projected separation of the 
outermost bound planet
at 5 Myr (black/solid lines), 10 Myr (blue/dotted lines), 
50 Myr (green/dashed lines), 100 Myr (red/dot-dashed lines)
and 500 Myr (magenta/dot-dot-dot-dashed lines).
\label{fsep}}
\end{figure}

Unfortunately, determining the semi-major
axes of planets on very wide orbits is difficult because the 
orbital timescale
greatly exceeds the typical time span of observations.  Instead,
direct imaging measures the star-planet separation projected onto the
plane of the sky.  In order to evaluate the detectability of such
planets, we present cumulative distributions for the projected
separation in Fig. \ref{fsep}.  In order to improve our statistics,
we calculate the projected separation for each of five times and ten
viewing geometries (by drawing the mean anomaly, pericenter angle, and
$\cos(i)$ from uniform random distributions).  For nearly circular
orbits, the planet-star separation remains close to the semi-major
axis throughout the orbit, and the projected separation averages to be
$\simeq0.7$ times the semi-major axis.  For highly eccentric orbits,
the planet spends most of its time near apocenter, so that the average
projected separation can exceed the semi-major axis, depending on the
orientation of the orbit relative to the line of sight.  We find that
the two effects approximately cancel for planets on wide orbits.
The cumulative distributions in Figs. \ref{fsemi}
and \ref{fsep} differ by just a few percent.

\begin{figure}

\plotone{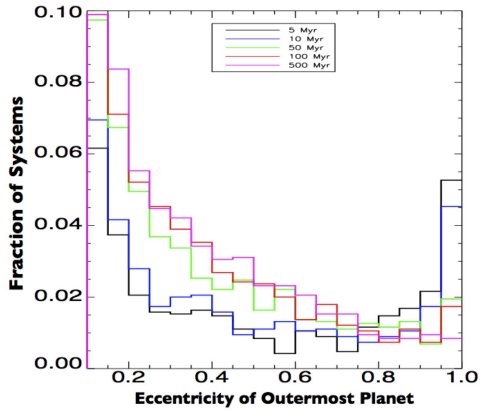}
\figcaption{
Histogram of the eccentricity of the outermost bound planet
at 5 Myr (black/solid lines), 10 Myr (blue/dotted lines), 
50 Myr (green/dashed lines), 100 Myr (red/dot-dashed lines)
and 500 Myr (magenta/dot-dot-dot-dashed lines).  Bound planets 
with nearly parabolic orbits are prevalent for several tens of Myr,
but eventually become unbound. 
\label{fecc}}
\end{figure}

%
Next, we investigate the eccentricities of the planets that have been
scattered onto wide orbits.  Previous studies have compared the
eccentricities of planets after planet scattering to radial velocity
observations.  For example, two-planet scattering rarely results
in an inner planet eccentricity greater than 0.8 following the
ejection of another planet \citep{forras2008}.  Similarly,
three-planet scattering results in only a modest fraction of inner
planets with eccentricities greater than 0.8 
\citep{chaetal2008}.  In this study, we focus instead on the 
eccentricities of planets that
have been scattered outwards.  Although many of these planets will
eventually become unbound, we consider the eccentricity distribution
of all planets still bound at each of several ages in Fig. \ref{fecc}.
The black/solid, blue/dotted, green/dashed, red/dot-dashed 
and magenta/dot-dot-dot-dashed lines on this histogram
correspond to 5, 10, 50, 100 and 500 Myr.  The last bin demonstrates that
the number of planets on nearly parabolic ($e > 0.95$) orbits decreases with
time, where the greatest difference is between the 10 and 50 Myr
curves.  Taken together, Figs. \ref{fsemi} and \ref{fecc} indicate
that planets on the verge of ejection typically have nearly linear
orbits, spending the vast majority of the time near apocenter, but
receiving a ``kick'' once each orbit when they briefly sweep into the
inner portion of the planetary system \citep{pansar2004}.
In our simulations, less massive planets are preferentially scattered into
wide orbits.  The majority of these planets are less than a Jupiter
mass, but only a few percent exceed $4 M_{Jup}$.


%


Figures \ref{fsemi}-\ref{fecc} do not include planets that have
already been ejected onto hyperbolic orbits ($e > 1$).  An unbound
planet will quickly become undetectable to a coronagraphic direct
imaging search with a relatively small field of view.  Assuming a
planet with an ejection velocity of $\sim GM_\star/a$, the ejected
planet would have traveled more than $10^5$AU away from the host
star (beyond $a_t$) in $\sim10^5$ years.  As a result, directly
detecting an ejected planet still in the vicinity
of its parent star is improbable.  However, such planets 
may still be detectable 
as free-floating planets by direct imaging of more distant star 
forming regions which image large fields of view.

\section{Detectability of Long Period Planets}\label{observe}
Here, we investigate the plausibility of detecting planets at large
separations with direct detection campaigns.  We take snapshots of
each planetary system from the previous section at several times: 5,
10, 50, 100 and 500 Myr, all times for which
\cite{baretal2008} provide theoretical models.  The first two
snapshots are particularly relevant for direct imaging of very young
stars, typically of star forming regions $\sim100$ pc from the Sun.
In those locations, models predict giants planets will be 
significantly more luminous in J
band than a blackbody of the same temperature.  The 100 and 500
million year snapshots are particularly relevant for observations of
nearby stars.  At these later ages, models predict that planets have
cooled enough that the J band luminosity is dramatically reduced, so L
band observations may become more favorable.

For direct imaging to detect planetary systems like our solar system, 
it is important to focus on the closest young stars 
(at $\approx 100$ pc), where the inner working 
angle corresponds to a small physical separation.  Most young stars in the 
immediate solar neighborhood are 
believed to have formed in small groups or associations, in which case the processes 
described in \S\ref{SecClusters} will not be relevant.  However, these same direct imaging 
campaigns may overlook young giant planets on wide orbits, as they lie beyond the 
detector's outer working angle.  When searching for planets at wide separations, it 
is possible or even advantageous to target more distant star forming regions.  Thus, 
it would be possible to search for planets that formed in more substantial 
clusters at a variety of ages.

\subsection{Detection Criteria}

\begin{figure*}
\hspace{0pt}
\centerline{
\psfig{figure=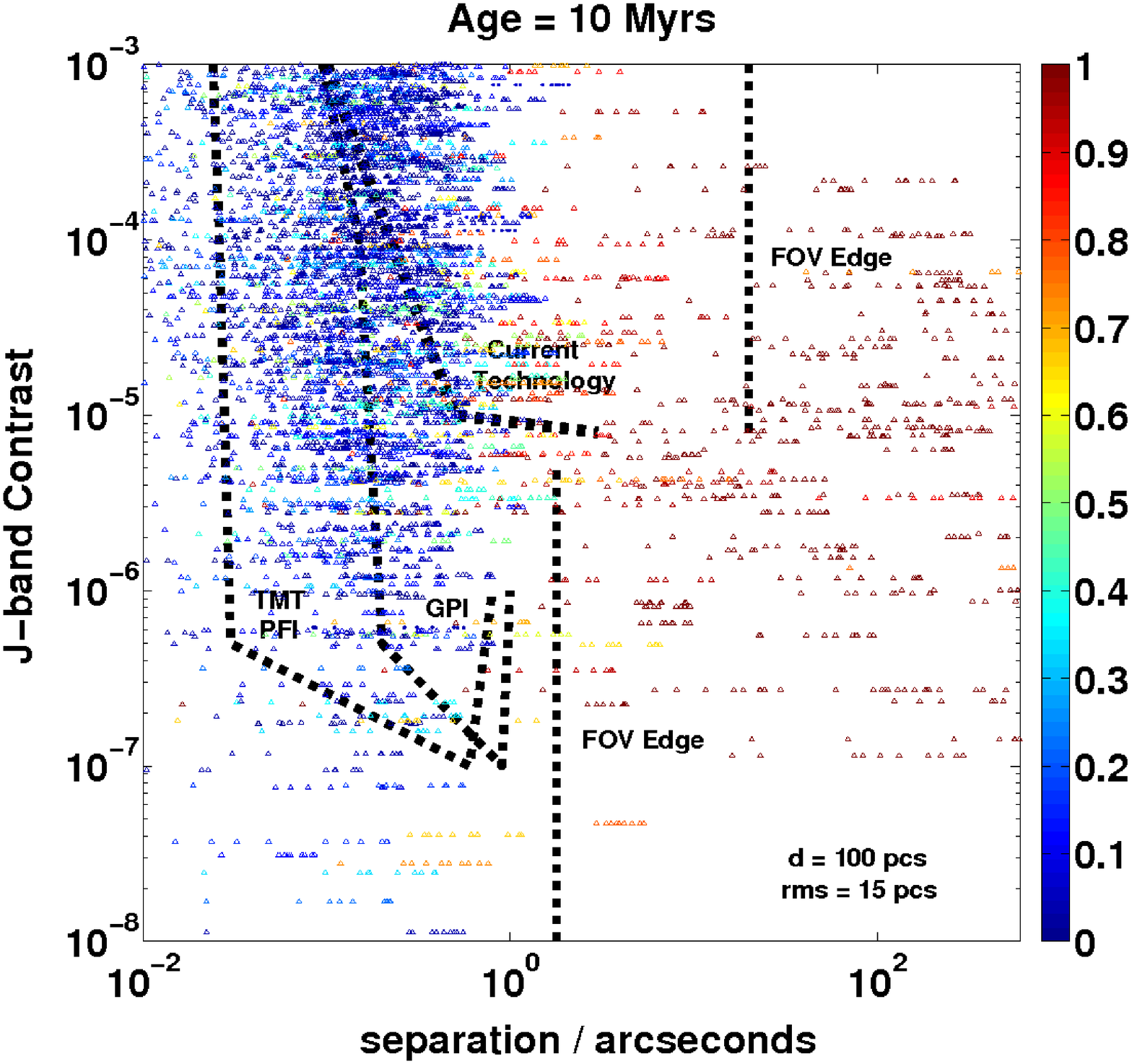,width=4.0truein,height=3.0truein}
\psfig{figure=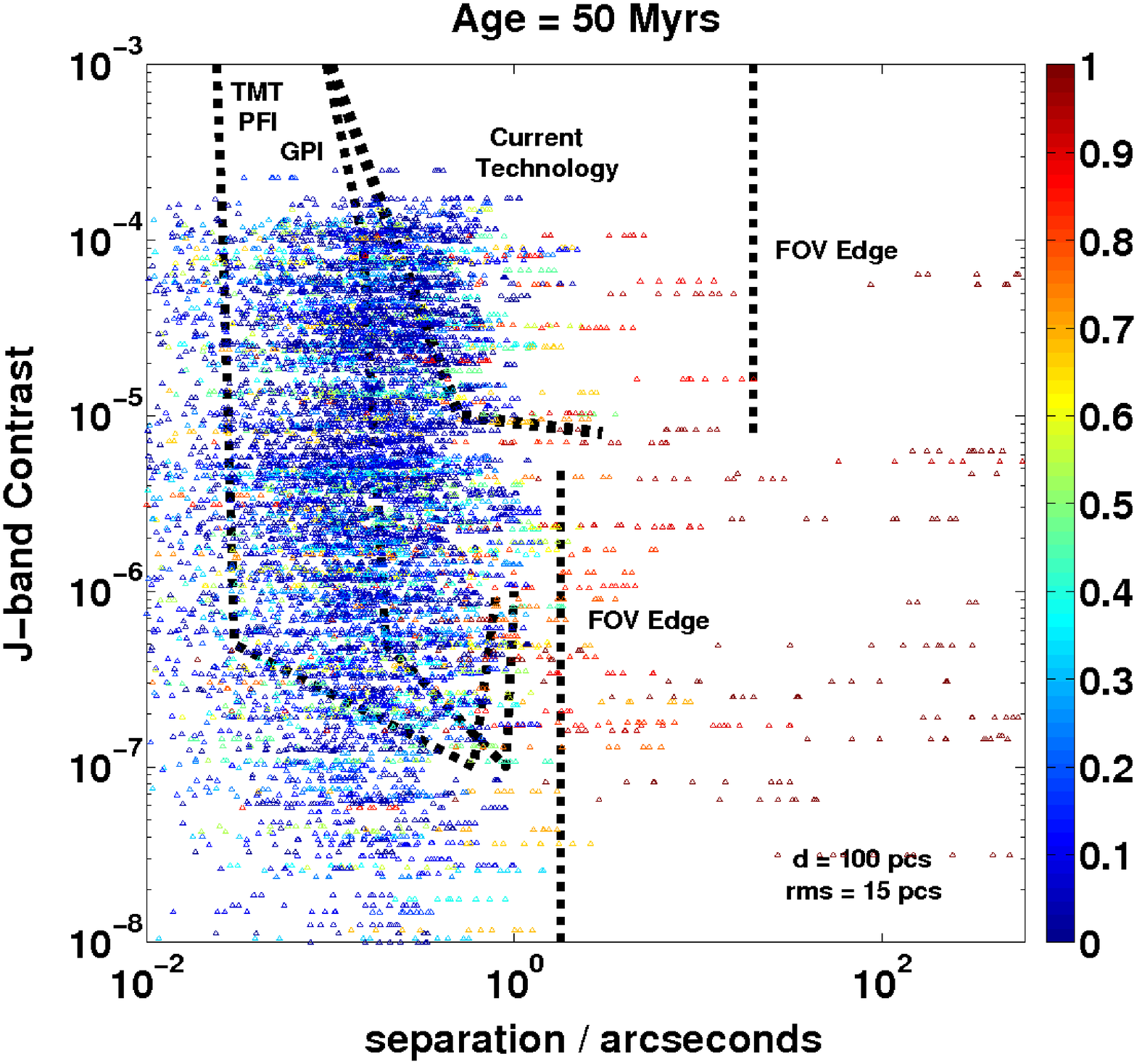,width=4.0truein,height=3.0truein}
}
\leftline{ }
\leftline{ }
\leftline{ }
\leftline{ }
\centerline{
\psfig{figure=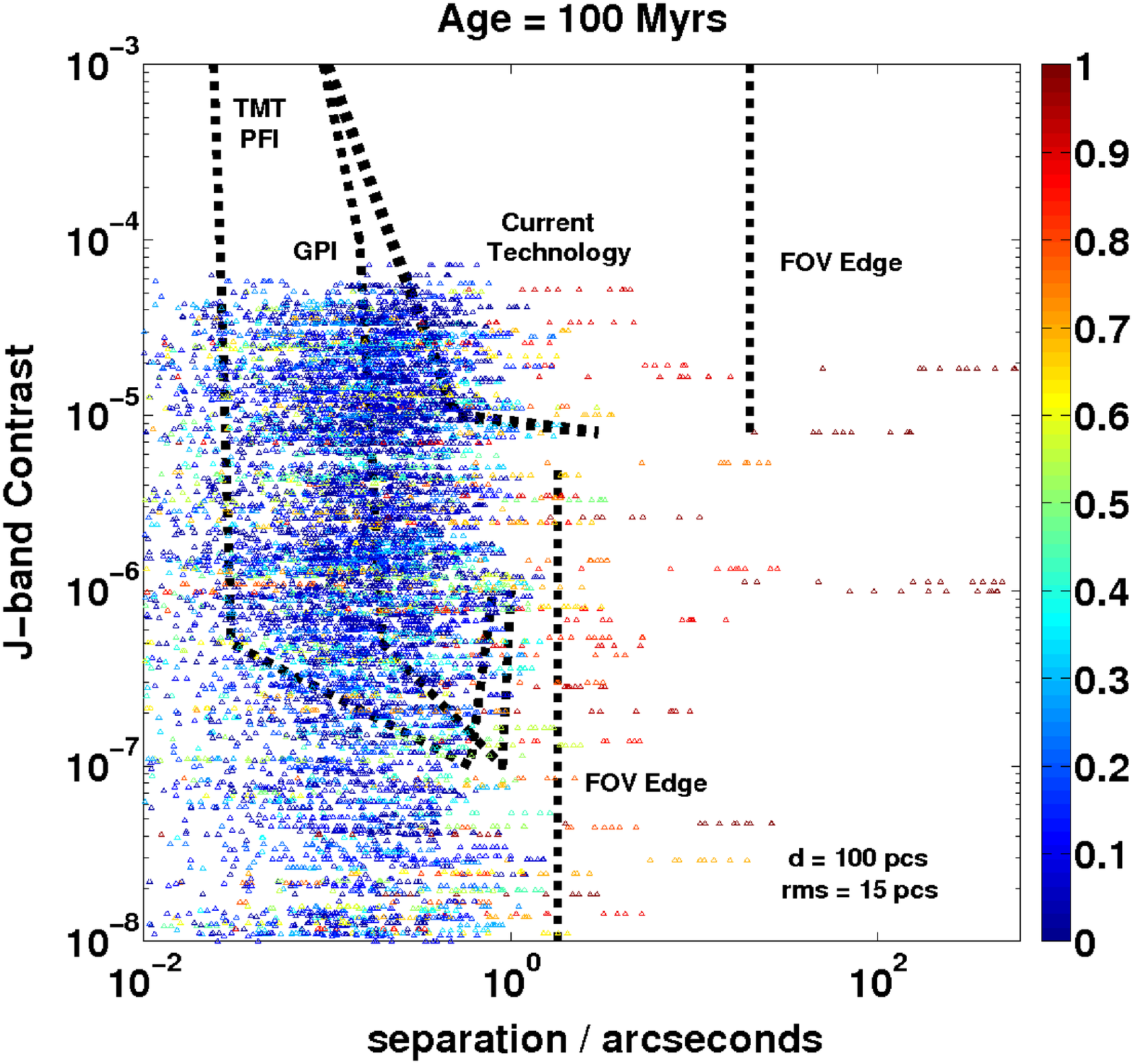,width=4.0truein,height=3.0truein}
\psfig{figure=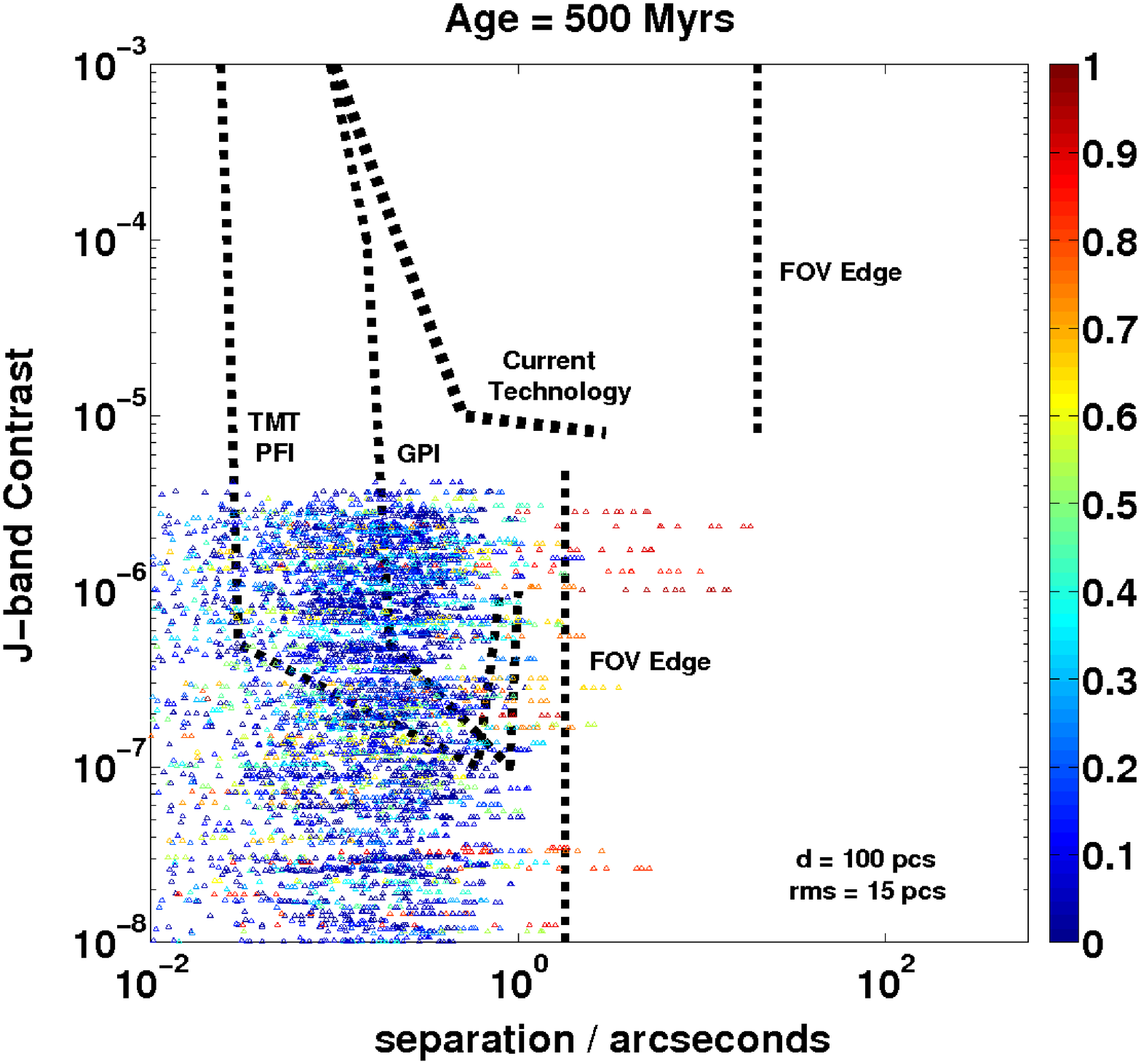,width=4.0truein,height=3.0truein}
}
\figcaption{
Contrast versus angular separation for a population
of exoplanets as a function of age. Data points are
color-coded/grayscaled according to eccentricity. Circles indicate that
all six planets in a given system remained gravitationally
bound. Triangles indicate that at least one of the planets
became unbound. Dashed lines indicate the approximate range
of detectability based on current and ground-based direct
imaging searches. Companions located exterior to the FOV edge are
not detectable. The estimates for GPI and the TMT PFI are somewhat
conservative, as they could achieve deeper effective
contrasts by a factor of order unity.
\label{fobs1}}
\end{figure*}

\begin{figure*}
\vspace{0pt}
\centerline{
\psfig{figure=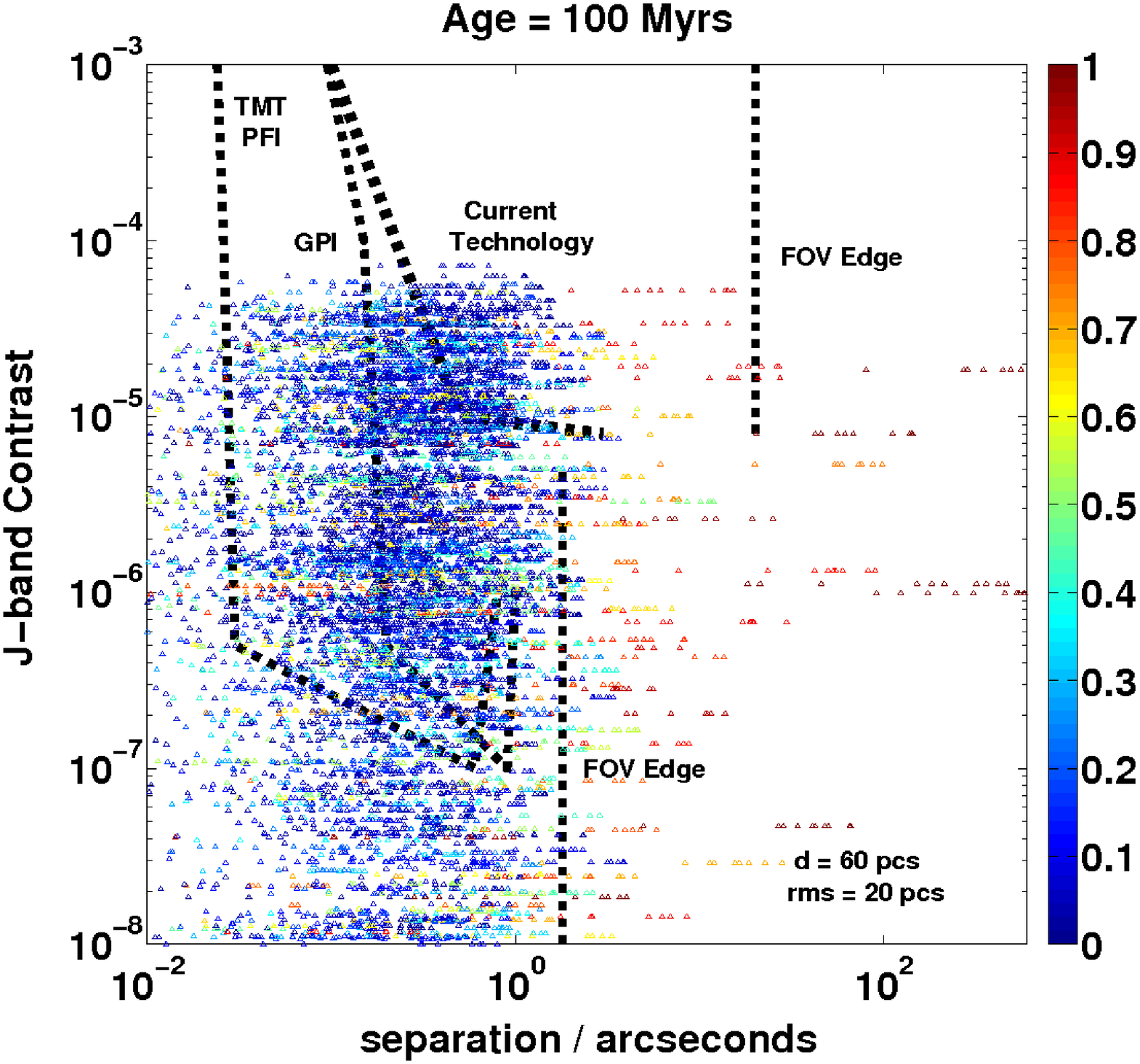,width=4.0truein,height=3.0truein}
\psfig{figure=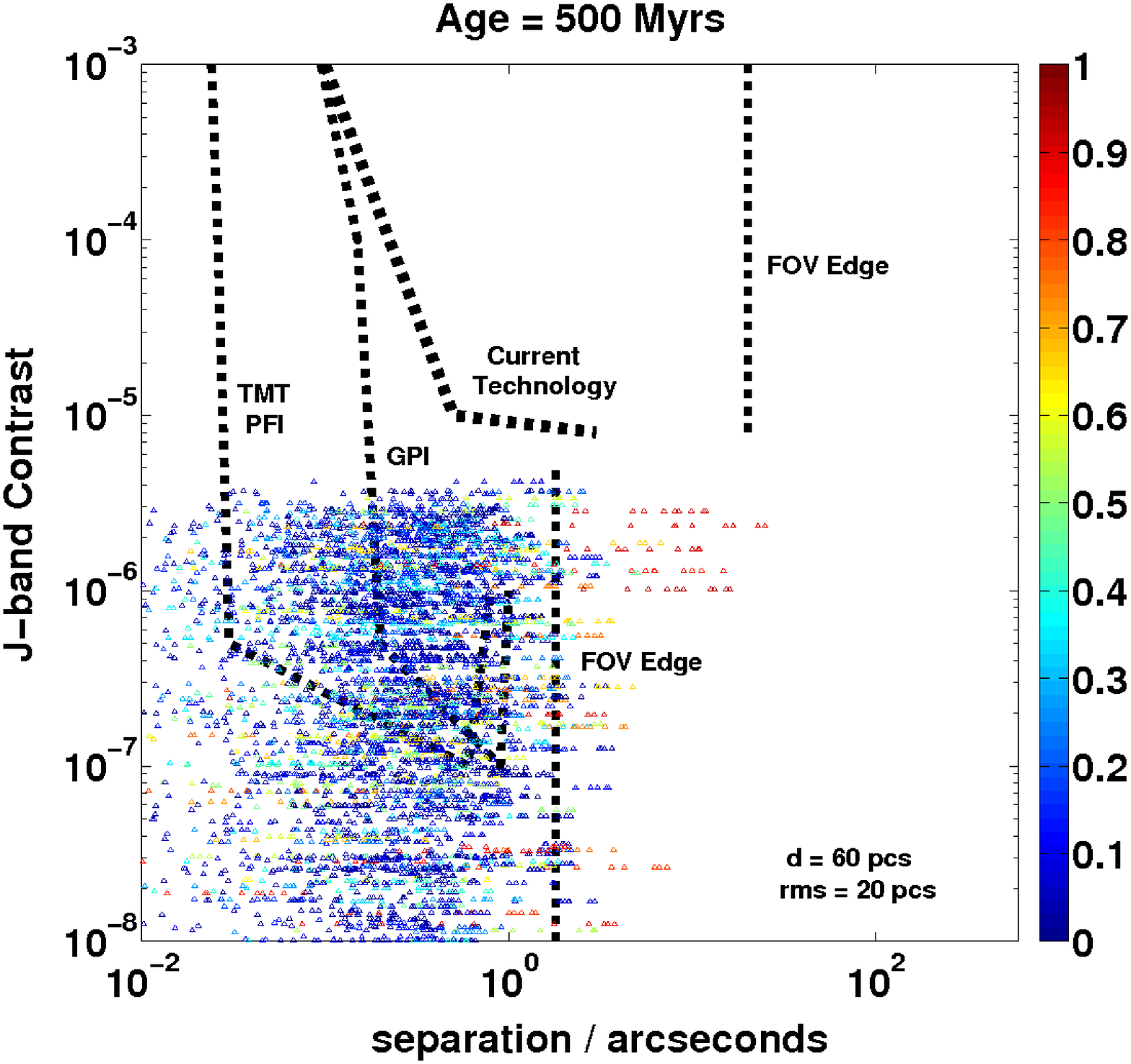,width=4.0truein,height=3.0truein}
}
\leftline{ }
\leftline{ }
\leftline{ }
\leftline{ }
\centerline{
\psfig{figure=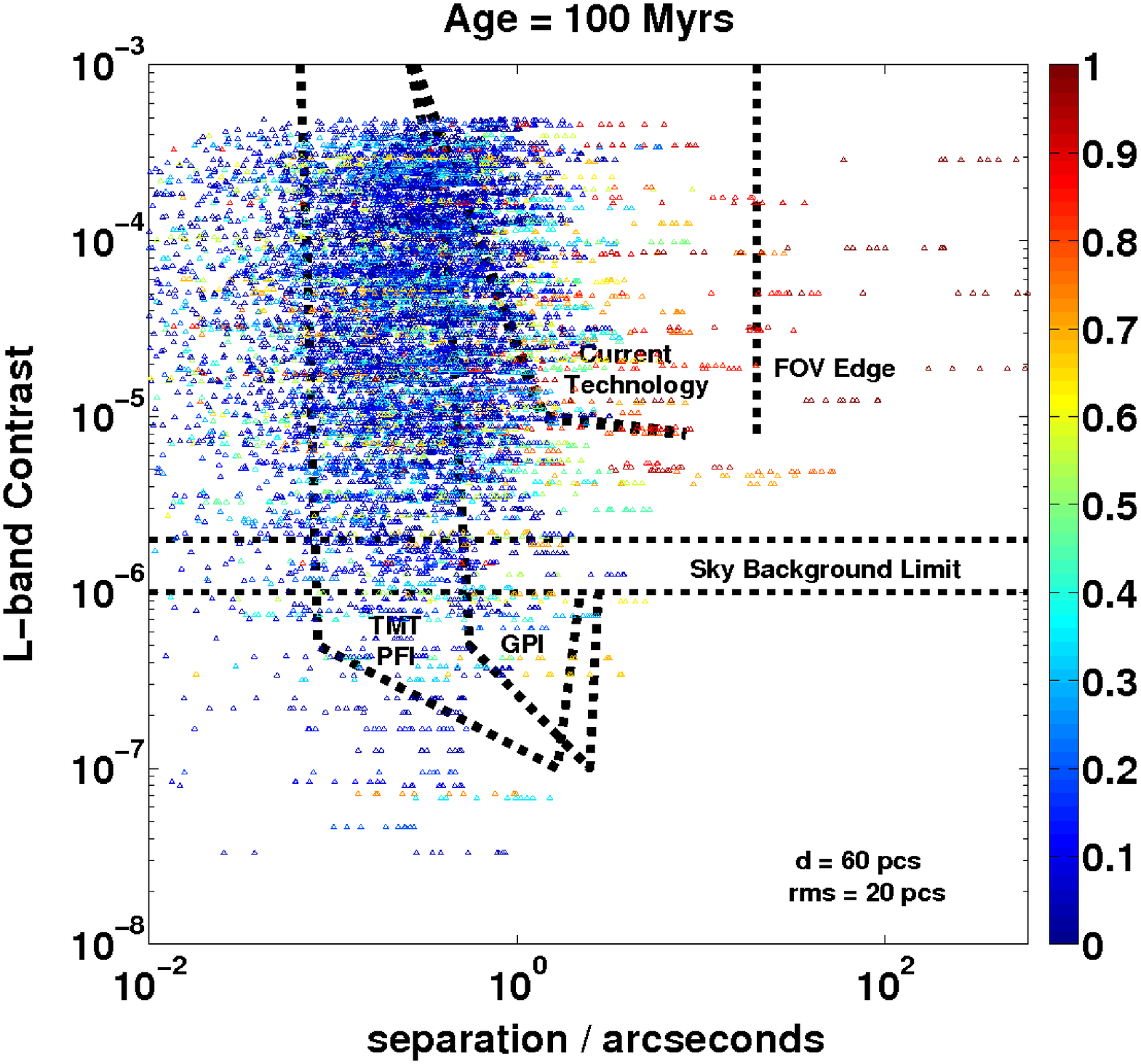,width=4.0truein,height=3.0truein}
\psfig{figure=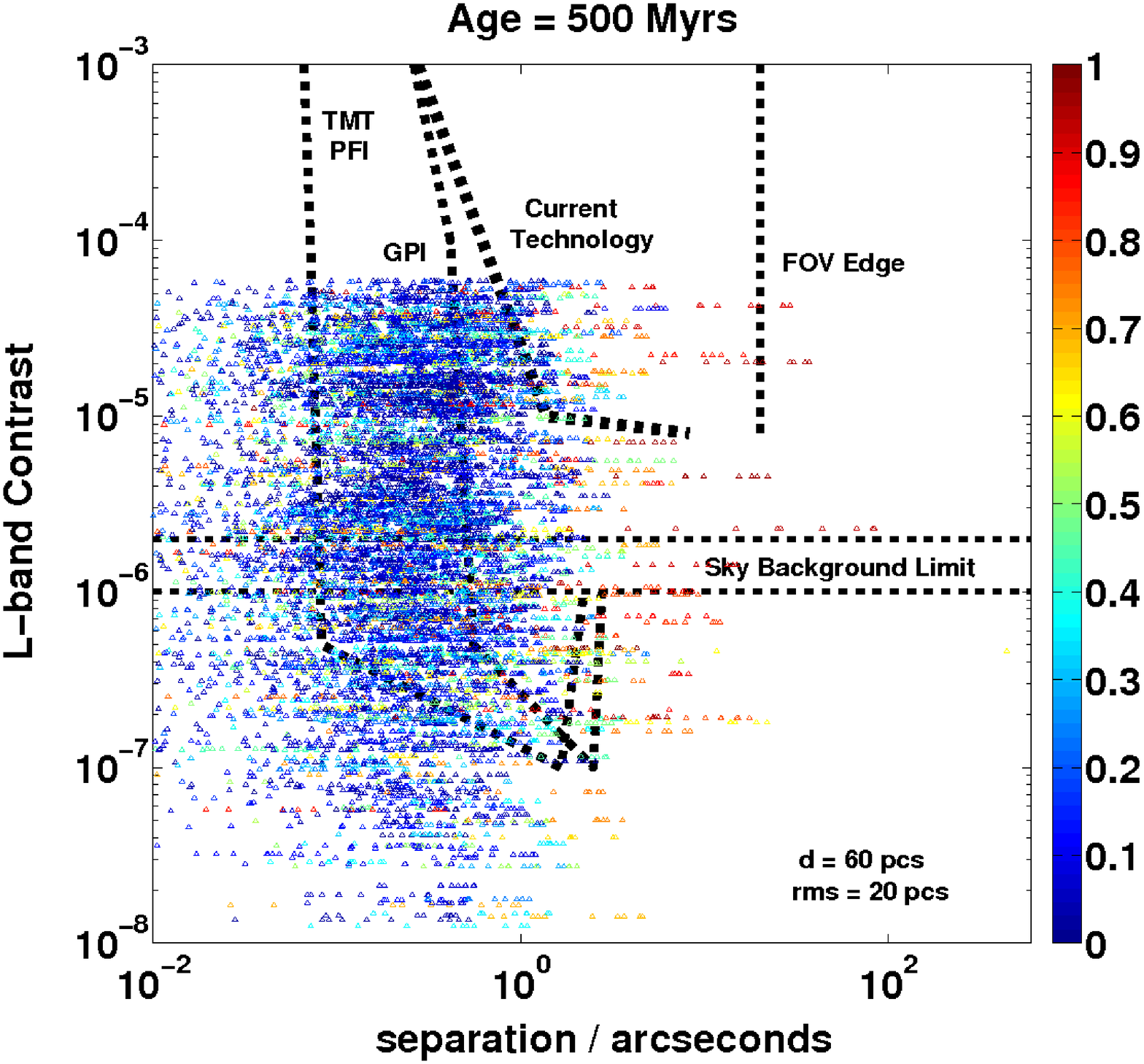,width=4.0truein,height=3.0truein}
}
\figcaption{
Contrast versus angular separation for a nearby
population of exoplanets in the J- ({\it upper panels})
and L ({\it lower panels})-bands.  Planets
are brighter in the L-band at older ages, but the sky background
is at least $10^{-6}$ times as bright as the brightest stars
{\em after} subtraction and places a hard-limit on ground-based
sensitivities. GPI is unlikely to conduct observations in the L-band.
\label{fobs2}}
\end{figure*}

Snapshots of each
planetary system from the previous section at 10,
50, 100 and 500 Myr, are plotted in terms of contrast and
angular separation in Fig. \ref{fobs1}. 
Contrast is calculated by comparing the absolute magnitude of the
planet to that of its host. We use the atmospheric models of 
\cite{girardi2002} and \cite{baretal2003} (and Baraffe et al. 2008,
private communication) to find the brightness of the stars and
exoplanets respectively. In Fig. \ref{fobs1}, we assume that a cluster of
stars, each with metalicity $Z=0.019$, is located at an average
distance of 100 pcs from the Sun with an RMS scatter of 15 pcs. In
Fig. \ref{fobs2}, we place the same stars at 60 pcs with an RMS scatter of 20
pcs. In each case, their positions within these limits are
randomized. For 
reference, we also include
approximate detection sensitivities from recent AO-coronagraph
surveys 
\citep[namely][]{biletal2007,kasetal2007,lafetal2007,nieetal2007}
as well as future instruments, such as GPI --
The Gemini Planet Imager \citep{macetal2006a} and the proposed
TMT PFI -- Thirty Meter Telescope Planet Formation Imager 
\citep{macetal2006b}.

Models predict that giant planets
will be several orders of magnitude more luminous in the near-IR
bands compared to a black-body of the same temperature at such young
ages.  The first two snapshots 
($10$+$50$ Myrs, top 2 panels of Fig. \ref{fobs1}) are 
particularly relevant for searching
members of star forming regions.  At these two times
we selected the J-band 
($\bar{\lambda}=1.25 \;\mu$m) because it
offers the brightest planets, best spatial resolution, and faintest
sky background in a regime where the atmosphere can still be
adequately controlled with AO.

The 100 and 500 Myrs snapshots are more relevant for observations of
nearby systems, such as Hyades or the Ursa Majoris cluster. At these
late stages, models predict that planets have cooled enough that the
J-band luminosity is dramatically reduced. In this case, L-band
observations, $\bar{\lambda}=3.8 \;\mu$m, may become favorable for
large angular separations \citep{heinze2007}. There is, however, a tradeoff,
because the L-band has the disadvantage of a bright and rapidly
fluctuating sky background \citep{phietal1999}, which ultimately
limits sensitivity (Fig. \ref{fobs2}). Chopping and nodding techniques can
only subtract out this source of noise (plus instrument and detector
noise) to one part in $\approx 10^5$, even when operating at the
photon noise limit. In the best case scenario, this translates to
contrast levels of $\approx 10^{-6}$, i.e. when targeting stars with
apparent magnitudes $L<4$. The prospects are better by nearly an
order of magnitude when conducting observations at the South Pole,
but constructing single dish telescopes
with sufficient angular resolution is currently difficult
given the harsh conditions.

It is also worth noting that the sensitivity of specialized
high-contrast imaging instruments is non-monotonic with angular
separation. This is due to the explicit removal of instrument
scattered light within the controllable spatial frequencies of the
AO system deformable mirror. The number of actuators and their
spacing defines this primary search area  -- the so-called
``dark-hole'' \citep{tratra2007}. GPI's field of view (FOV) is
3.6". Current high resolution instruments have somewhat larger FOV's
\citep[e.g.][]{hayetal2001}.

\begin{figure}

\plotone{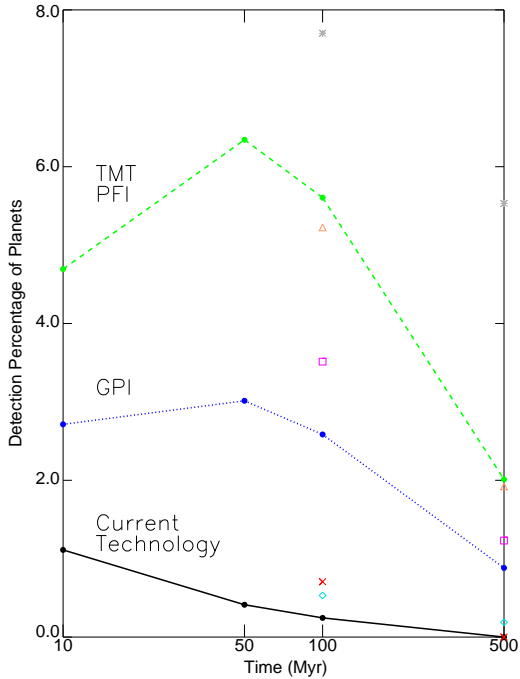}
\figcaption{
Percent of all planets from an enhanced distribution 
(where 10 viewing 
geometries for each initially 6-planet system are included)
which are detectable within the bounds imposed in Figs. 
\ref{fobs1} and \ref{fobs2} according to current 
technology (solid/black curve), GPI (blue/dotted curve)
and TMT PFI (green/dashed curve) in the J-band for a 
cluster located at an average
distance of 100 pcs from the Sun with an RMS scatter of 15 pcs. 
All isolated symbols refer to a cluster located at an average
distance of 60 pcs from the Sun with an RMS scatter of 20 pcs
and represent: J-band, currently detectable (red crosses); 
J-band, GPI detectable (magenta squares); 
J-band TMT PFI detectable (salmon triangles); 
L-band currently detectable (aqua diamonds);
and L-band TMT PFI detectable (grey asterisks).
\label{fstat}}
\end{figure}

Finally, in Fig. \ref{fstat}, we plot the fraction 
of orbit realizations of all planets which would 
be detectable according to Figs. \ref{fobs1} 
and \ref{fobs2} as a function of age,
band, and separation.  We emphasize that this 
fraction is representative of all planets in dynamically
active systems (those which undergo instability),
according to our initial conditions.
Current technology should allow for the direct
imaging detection of planets at ages $\lesssim 100$ Myr, whereas
GPI and TMT PFI are predicted to discover several times more
planets.  Both instruments are likely to discover the greatest number
of planets at an age of $\approx 50$ Myr.  Importantly,
current J-band (red crosses) and L-band (aqua diamonds) observations
demonstrate the slight improvement the latter affords at later times; 
the projected TMT PFI observations in the J-band (salmon triangles)
and L-band (grey asterisks) demonstrate the marked improvement
the L-band provides at both $100$ Myr and $500$ Myr.


\section{Discussion}
We have shown that planets could survive in very-wide orbits.  For the
relatively young stars typically surveyed by direct imaging campaigns,
planets could survive at distances of up to the tidal limit.  As for a
potential formation mechanism, we consider the role of planet-planet
scattering, with a focus on the implications for long-period planets.
Because the process of planet-planet
scattering typically extends for tens of millions of years, most stars
will not remain in a high-density stellar environment long enough for
close stellar encounters to be effective at removing long-period
planets (with $a < a_t, a_{\mathrm{ps}}$; see Eqs. \ref{at} and \ref{aps}).

We performed direct n-body simulations of dynamically active
planetary systems, including the effects of
galactic tides, and show the planet-scattering can indeed form
long-period planets.  Of particular interest, we found that at ages of
$\simeq~10-50$ Myr, there is a sizable population of planets
on wide and highly eccentric orbits.  Although most of these planets will
eventually be ejected, they can persist on wide orbits for several
tens of millions of years.  We caution that our models predict a
significant temporal evolution of this population.  Therefore,
statistical analyses of direct imaging surveys should account for the
possibility of a time-variable population of planets on wide orbits.

We considered the detectability of the planets scattered into
long-period orbits in our simulations.  Planet models are
highly uncertain, largely due to the uncertainty in initial conditions
\citep[e.g.][]{maretal2007}.  Nevertheless, we show that even 
assuming ``cold-start''
models, direct imaging surveys are already sensitive to giant planets
in nearby star forming regions.  Future direct imaging surveys will
become sensitive to both closer planets and planets with smaller
luminosities.

\begin{figure}

\plotone{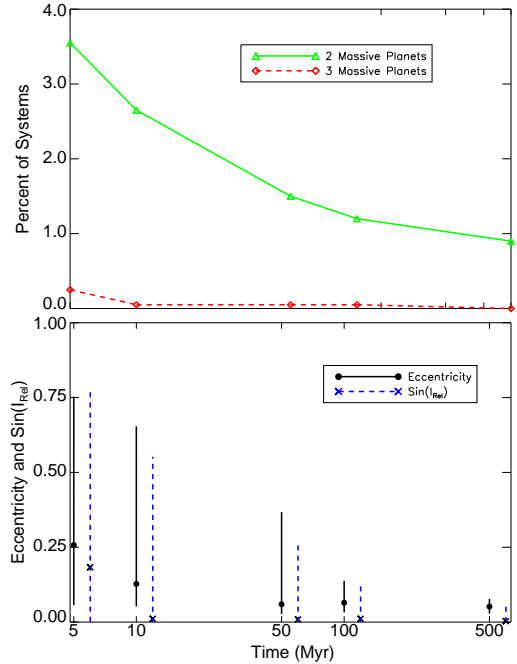}
\figcaption{
Statistics of systems with multiple bound planets 
all on wide orbits (projected separation $\ge 40$ AU) 
as a function of age.  The upper panel plots the fraction 
of all systems with two (black/solid line and dots), three 
(blue/dotted line and crosses) and four (green/dashed line 
and squares) planets remaining beyond 40 AU.  The red/dot-dashed line with 
triangles and the magenta/dot-dot-dot-dashed line with diamonds 
refer to systems with two or three massive ($\ge 0.3 M_{J}$) planets 
remaining.  The lower panel showcases the median eccentricities 
(dots and black/solid lines) and sine of the relative inclinations (crosses 
and blue/dashed lines) of all systems with two planets on wide 
orbits, and their 25-75 percentile ranges.  The blue/dashed lines 
are slightly offset from the other lines for clarity.
\label{multfig}}
\end{figure}

Recent interest in the candidate three-planet HR 8799 system 
\citep{maretal2008} motivates us to consider the multiplicity of planets
from the same system all in bound wide orbits.  The properties (such
as eccentricity and inclination) of the orbits of those three planets
are largely unconstrained at the present time.  
The planet masses are approximated to lie in the $5$ −- $13 M_{Jup}$ 
range and the projected separations are estimated to be 24, 38, 
and 68 AU (Marois et al. 2008).  In our simulations of \S3, 
the initial semi-major axis of the outermost planet is 
typically 15-30 AU, so any planets beyond 40 AU have undergone 
significant scattering.  In order to evaluate the frequency 
of multiple planets at wide separations due to planet 
scattering, we tally how many systems contain multiple 
planets on bound orbits with semi-major axes exceeding 40 AU
and masses exceeding $5 \times 10^{-4} M_{\odot}$. 
At (5, 10, 50, 100, 500) Myr, 
(71, 53, 30, 24, 18) [76, 59, 34, 26, 19] systems 
contain two such planets and 
(5, 1, 1, 1, 0) [7, 1, 1, 1, 0] systems 
contain three or more such planets (see Fig \ref{multfig}).

\begin{figure}

\plotone{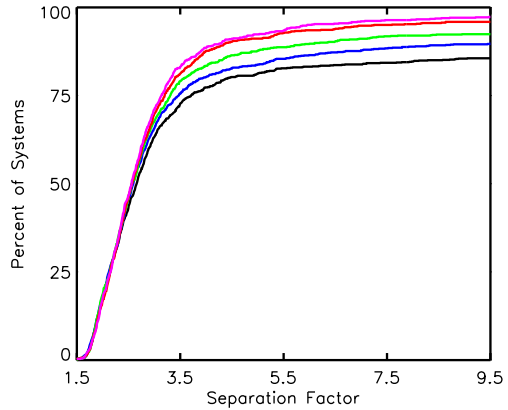}
\figcaption{
Mutual separations of outermost 3 planets in each system,
such that each adjacent pair of the 
outermost 3 planets has a projected separation within a 
factor of the values on the X-axis.  Black/solid, blue/dotted, green/dashed, 
red/dot-dashed and magenta/dot-dot-dot-dashed lines refer to 
snapshots at 30, 40, 50, 100, and 150 Myr.
\label{fplsep}}
\end{figure}

These results suggest that 
3.8\% (0.9\%) of dynamically active planetary systems could 
contain multiple massive planets on wide orbits at ages of 
5 Myr (500 Myr and beyond).  Further, 66\% (40\%) 
of systems containing multiple planets on wide orbits 
with ages less than 10 Myr (100 Myr) will become 
unstable on longer timescales.  This model predicts that 
multiple planets on wide orbits will become less common 
with increasing stellar age (Fig. \ref{multfig}).  Figure
\ref{fplsep} also demonstrates this trend by displaying
the percent of systems with each adjacent pair of the 
outermost 3 planets having a projected separation within
factors appearing on the X-axis.  Black, blue, green, 
red and magenta curves refer to data 
at 30, 40, 50, 100, and 150 Myr, ages chosen
to correspond to the likely age of the three planets in
HR 8799 \citep{maretal2008}.   On this plot, $< 1\%$ of all 
our systems simulated would be as ``packed" as HR 8799 
(with an X value of $38/24 = 1.58$), suggesting that the
planets in that system are unlikely to have been scattered
to those locations without resonant locking \citep{fabmur2008} 
or migrating through a disk.

Young 
systems containing multiple planets at wide separations 
would be expected to have sizable eccentricities, even larger 
than average of extrasolar planets.  With time, 2-planet 
systems with large eccentricities tend to be destroyed, so the average 
eccentricity decreases to a more modest value ($0.14$), 
very similar to that observed for multiple planet systems 
discovered by radial velocity searches \citep{wrietal2008}.  
The average relative inclination between two planets on 
wide orbits also decreases with time from $28.4\%$ to $9.5\%$ 
degrees between 10 and 500 Myr.  These results can be 
used to inform the interpretation and dynamical modeling 
of multiple planet systems found by direct 
imaging \citep{fabmur2008}.

In the future, comparing the frequency of long-period planets as a
function of stellar age could provide a significant constraint on
planet formation models.  Such observations could determine the
timescale for the onset of strong instabilities in multiple planet
systems and place constraints on the typical number of giant planets
formed (which may be greater than the number of giant planets around
mature stars).  Such constraints would be particularly interesting,
when combined with constraints on free-floating planets from direct
imaging of star forming regions and/or gravitational microlensing
observations.  

In our simulations excluding the galactic tide, 40\% [$42$\%] of 
all planets (totaling $13.6$\% [$13.7$\%] of the planetary mass) 
were ejected.  Of the planets ejected, $21.2$\% ($28.9$\%)
[$21.0$\% ($28.9$\%)] had 
a mass (mass ratio) greater 
than $0.3 M_{Jup}$ ($0.3M_{Jup}/M_{\odot}$) and 
$54.1$\% ($60.1$\%) [$54.2$\% ($60.1$\%)] 
of planets had a mass 
(mass ratio) between $5 M_{\oplus}$ ($5M_{\oplus}/M_{\odot}$) and 
$0.3 M_{Jup}$ ($0.3 M_{Jup}/M_{\odot}$).
The simulations which included the effects from the galactic tide 
yielded nearly identical results, varying the above values by
at most by a few percent.  All these ejected planets would become 
free-floating planets.  After they fade and become undetectable 
by direct imaging, 
the best chance for detecting this population of planets is via 
gravitational microlensing.  Although the optical depth for microlensing is 
independent of the mass function of free-floating planets, the event rate and 
event timescale are sensitive to the mass function.  The event
rate for ejected microlensing planets which are more massive than
$0.3 M_{Jup}$ ($0.3 M_{Jup}/M_{\odot}$) is 
$56.6\%$ ($64.4$\%)                   
[$56.2\%$ ($64.4$\%)] of
the total event rate for all unbound masses, but
other mass bins will produce different event rates. 
In order
to better quantify this dependence on mass, we bin the masses
uniformly in $\log{M}$ and for sufficiently high sampling,
estimate that a power-law index
of $[\approx 0.37]$ approximates the positive correlation
between event rate (as the dependent variable) and mass.

In reality, there would be additional microlensing events due to 
1) planets still bound to their parent stars, and 2) small planets or 
protoplanets there were not included in our simulations.
Therefore, the above numbers should be interpreted as a lower limit.
Regarding the contribution from bound planets, $38.0$\% [$39.9$\%] of all 
the initial planets (totaling $81.5$\% [$81.3$\%] of the planetary 
mass) remain bound.
Of these, $76.5$\% ($84.1$\%)  [$76.2$\%  ($84.1$\%)] had a 
mass (mass ratio) greater
than $0.3 M_{Jup}$ ($0.3M_{Jup}/M_{\odot}$), and accounted for
$93.8\%$ ($96.4\%$) [$93.6$\% ($96.4$\%)] of the overall microlensing event rate.
Unfortunately, 
existing surveys are prone to missing short-duration events and hence biased 
towards finding more massive planets (especially for free-floating planets 
where there is not a longer duration stellar microlensing event to trigger 
intensive observations).  These results suggest that a microlensing survey 
that efficiently detected the short events due to free-floating terrestrial 
planets would provide a useful diagnostic for the rate at which terrestrial 
planets form in dynamically active systems (relative to the rate of giant 
planet formation in similar systems).  When combined 
with statistical information 
about the frequency of terrestrial and giant planets in 
mature planetary systems, this synergy would provide a 
probe of the chaotic phase of planet formation and a test of 
the planet-planet scattering model for exciting 
eccentricities of extrasolar planets.

We considered the detectability of the planets scattered into
long-period orbits in our simulations.  Indeed, planet models are
highly uncertain, largely due to the uncertainty in initial conditions
\citep[e.g.][]{maretal2007}.  We show that even assuming ``cold-start''
models, direct imaging surveys are already sensitive to giant planets
in nearby star forming regions.  Future direct imaging surveys will
become sensitive to both closer planets and planets with smaller
luminosities.

\acknowledgments{We thank the anonymous referee for his
keen observations, and Daniel Apai, John Chambers,
Vacheslav Emel'yanenko, Daniel Fabrycky, Elizabeth Lada, 
Eduardo Martin, Michael Meyer, Ruth Murray-Clay,
Steinn Sigurdsson, Jonathan Tan,
and Scott Tremaine for sharing their insights.
The authors acknowledge the University of Florida High-Performance Computing 
Center for providing computational resources and support that have contributed 
to the research results reported within this paper.
}

\pagebreak


\pagebreak


\pagebreak



\pagebreak

%

\pagebreak



\pagebreak



\pagebreak

\pagebreak

\pagebreak

%

\pagebreak



\pagebreak

\end{document}